\documentclass[a4paper]{article}
\usepackage{INTERSPEECH2020}
\usepackage{url}
\usepackage{graphicx}
\usepackage{caption}
\usepackage{subfigure}

\title{Hybrid Transformer/CTC Networks for Hardware Efficient Voice Triggering}
\name{Saurabh Adya, Vineet Garg, Siddharth Sigtia, Pramod Simha, Chandra Dhir}
\address{Apple}
\email{(sadya,vineetgarg,sidsigtia,psimha,cdhir)@apple.com}

\begin{document}

\maketitle
\begin{abstract}
We consider the design of two-pass voice trigger detection systems. We focus on the networks in the second pass that are used to re-score candidate segments obtained from the first-pass. Our baseline is an acoustic model(AM), with BiLSTM layers, trained by minimizing the CTC loss. We replace the BiLSTM layers with self-attention layers. Results on internal evaluation sets show that self-attention networks yield better accuracy while requiring fewer parameters. We add an auto-regressive decoder network on top of the self-attention layers and jointly minimize the CTC loss on the encoder and the cross-entropy loss on the decoder. This design yields further improvements over the baseline. We retrain all the models above in a multi-task learning(MTL) setting, where one branch of a shared network is trained as an AM, while the second branch classifies the whole sequence to be true-trigger or not. Results demonstrate that networks with self-attention layers yield $\sim$60\% relative reduction in false reject rates for a given false-alarm rate, while requiring 10\% fewer parameters. When trained in the MTL setup, self-attention networks yield further accuracy improvements. On-device measurements show that we observe 70\% relative reduction in inference time. Additionally, the proposed network architectures are $\sim$5X faster to train.

\end{abstract}

\section{Introduction}

\label{sec:intro}

There are a growing number of devices with speech as the primary means of user input for e.g. smart speakers, headphones and watches. As a result, voice trigger detection systems have become an important component of the user interaction pipeline as they signal the start of an interaction between the user and a device. Since these systems are deployed entirely on-device, there are several considerations like privacy, latency, accuracy and battery/power consumption that inform their design. We employ a two-stage architecture for the trigger detectors \cite{MLBlogHS,gruenstein2017cascade,wu2018monophone}, where a low-power first-pass detector receives streaming input from the microphone and is always running \cite{sigtia2018vt}. If a detection is made at this stage, larger more complex models are used to re-score the candidate acoustic segments from the first-pass \cite{sigtia2020mtl}. This design offers a balance between power/battery consumption which is determined by the first-pass and overall accuracy which is determined by the larger models in the second-pass. 

This paper aims to improve the architecture of the second-pass detectors in order to make better use of the available on-device hardware. Recent approaches to this problem have explored a number of neural network architectures like DNNs \cite{chen2014small,choi2019temporal}, CNNs \cite{sainath2015convolutional,arik2017convolutional,kao2019sub} and RNNs \cite{fernandez2007application,he2017streaming,yamamoto2019small,sigtia2020mtl}. Here we experiment with using stacks of self-attention layers \cite{cheng2016long,lin2017structured,vaswani2017attention} to replace bidirectional LSTM (BiLSTM) layers. This design is motivated by 2 observations.  Firstly, the second-pass models receive the entire input audio for re-scoring at once and do not need to be run in a streaming setting. Previously \cite{sigtia2018vt} we took advantage of this fact by using BiLSTM layers to read the input from both directions. However this arrangement requires sequential computations at every layer in the network, which can be slow. Self-attention layers, on the other hand, process the entire input sequence with feed-forward matrix multiplications (c.f. Section \ref{sec:background}). Secondly, we can improve training and inference times significantly, because the feed-forward computations in the self-attention layers with large matrix multiplication operations can be easily parallelized using the available hardware. 

In previous work \cite{sigtia2020mtl}, we argued that there are two natural ways to design a second-pass voice trigger detector. The first method is to train a monophone AM and use this model to compute the probability for the phone sequence in a trigger phrase given an acoustic segment from the first-pass. The second method is to directly train a binary classifier to discriminate between true examples of the trigger phrase and the false examples including easily confusable/phonetically similar examples. The first method has the advantage that we can use large transcribed training sets to train the main speech recognizer for a given language, but suffers from the fact that phonetically confusable utterances are assigned similar scores. The second method has the advantage that we train using exactly the correct objective function for the task at hand. However, collecting large training sets for this discriminative task in a privacy preserving way is extremely challenging. We proposed to combine the useful properties of both approaches using multi-task learning (MTL) and observed significant improvement in accuracies. In the present work, we build on these ideas by replacing the stack of BiLSTM layers with stacks of self-attention layers in order to better utilize the on-device hardware during inference. Our results show that the self-attention networks yield similar accuracies to the models in \cite{sigtia2020mtl}, \emph{without} requiring the additional discriminative training data and requiring 10\% \emph{fewer} model parameters. This result is a significant improvement as it allows us to train more accurate models for languages where we only have access to general AM training data but do not have a dataset of true triggers and false alarms. We also show that adding discriminative training data to these networks yields further improvements, significantly improving over the baselines presented in \cite{sigtia2020mtl}. Finally, we measure the inference speed of the proposed models on some recent hardware devices and we find that inference time with the proposed networks can be reduced by upto 70\%.

\section{Model Architectures}

\label{sec:background}

In this section we present details of the baseline architecture and the proposed modifications to the baseline. To recap, our motivation is two-fold; to improve the accuracy of the models and to make better use of the available on-device hardware. 
We start with replacing the BiLSTM layers in the baseline with self-attention layers. We find that this modification yields better accuracies on 2 evaluation sets while requiring fewer parameters. Next, we add an auto-regressive decoder as an additional/auxiliary loss (Figure 1). We find that jointly minimizing the connectionist temporal classification (CTC) loss and the cross-entropy loss yields further improvements compared to minimizing only the CTC loss. Note that during inference, we only use the encoder part of the network (Figure 1), to avoid sequential computations in the auto-regressive decoder. Therefore the transformer decoder can be seen as regularizing the CTC loss. Alternatively, this setup can be viewed as an instance of multi-task learning where we jointly minimize 2 different losses. %Note that we do not include the parameters in the decoder in our parameter counts since it is not used at inference. 

%As we explained briefly in Section \ref{sec:intro}, we explore replacing the BiLSTM layers with self-attention layers. In this context, we also introduce a novel loss function to the Transformer \cite{vaswani2017attention} architecture in which we combine the traditional cross-entropy loss on its decoder along with CTC \cite{ctc_graves} loss on its encoder. Similar approach has been explored with a BiLSTM network in \cite{watanabe-hybrid} \cite{Xiao-isclp2019}. The self-attention layers replace the sequential computations in the BiLSTM layers with feed-forward computations which improves both training and inference times because of better parallelization capabilities of modern hardware. A full encoder-decoder transformer architecture is a sequence-to-sequence model and is trained by minimizing cross-entropy (CE) loss at the decoder output. We did not explore the full encoder-decoder transformer architecture because they are particularly slow during inference due to recurrent nature of decode in attention decoder. In summary, we experimented with 2 new architectures: (1) encoder only architecture which replaces BiLSTM layers with the self-attention layers and is trained with CTC loss (2) encoder-decoder transformer architecture trained to jointly minimize the encoder CTC loss and the decoder CE loss. During inference, we remove the attention decoder branch and instead use a CTC decoder on encoder per frame outputs. We describe both these architectures along with our baseline BiLSTM model below.

\subsection{Baseline LSTM + CTC}
\label{sec:bilstm}
We use the same baseline BiLSTM architecture as in \cite{sigtia2020mtl}. We compute 40-dimensional mel-filterbank features from the audio at 100 frames-per-second (FPS). At every time-step we splice 7 frames together to form a 280-dimensional input window, and we subsample the sequence of windows by a factor of 3. The inputs are presented to a stack of 4 BiLSTM layers with 256 units each, resulting in 5.4 million trainable weights. The output layer comprises an affine transformation followed by the softmax non-linearity resulting in 54 outputs which span the set of context-independent phones (monophones) and sentence and word boundaries. The network is trained by minimizing the CTC loss given a large training set containing pairs of speech utterances and their corresponding text transcriptions. 

\subsection{Self-attention + CTC}
\label{sec:attention_ctc}
Next, we replace the BiLSTM layers with a stack of self-attention layers \cite{vaswani2017attention}. We process the inputs same as before but we add a 280-dimensional \emph{fixed} positional encoding to each input frame. We use exactly the same positional encoding scheme of alternating sine and cosine waves with varying wavelengths as proposed in \cite{vaswani2017attention}. We use a stack of 6 self-attention layers and the computation performed by each layer is depicted in Figure \ref{fig:model_arch}. We use 4 heads for each of the self-attention transforms with each head yielding a 64-dimensional key, query and value vectors. Each head is concatenated to output a 256-dimensional vector. We use a hidden size of 1024 dimensions for the feed-forward layer. We also use skip connections for both the self-attention and the feed-forward layers and we apply Layer Normalization \cite{ba2016layer} to the outputs of both transforms. The resulting network contains 4.8 million trainable weights, which is 10\% smaller than the BiLSTM baseline. We train the network by minimizing the CTC loss as before, using the same training data. We refer to this configuration as \emph{Self Attention Encoder} in the rest of the paper. 

\subsection{Self-attention + CTC + Decoder}
\label{sec:attention_ctc_decoder}

\begin{figure}
\centering
\includegraphics[width=\linewidth]{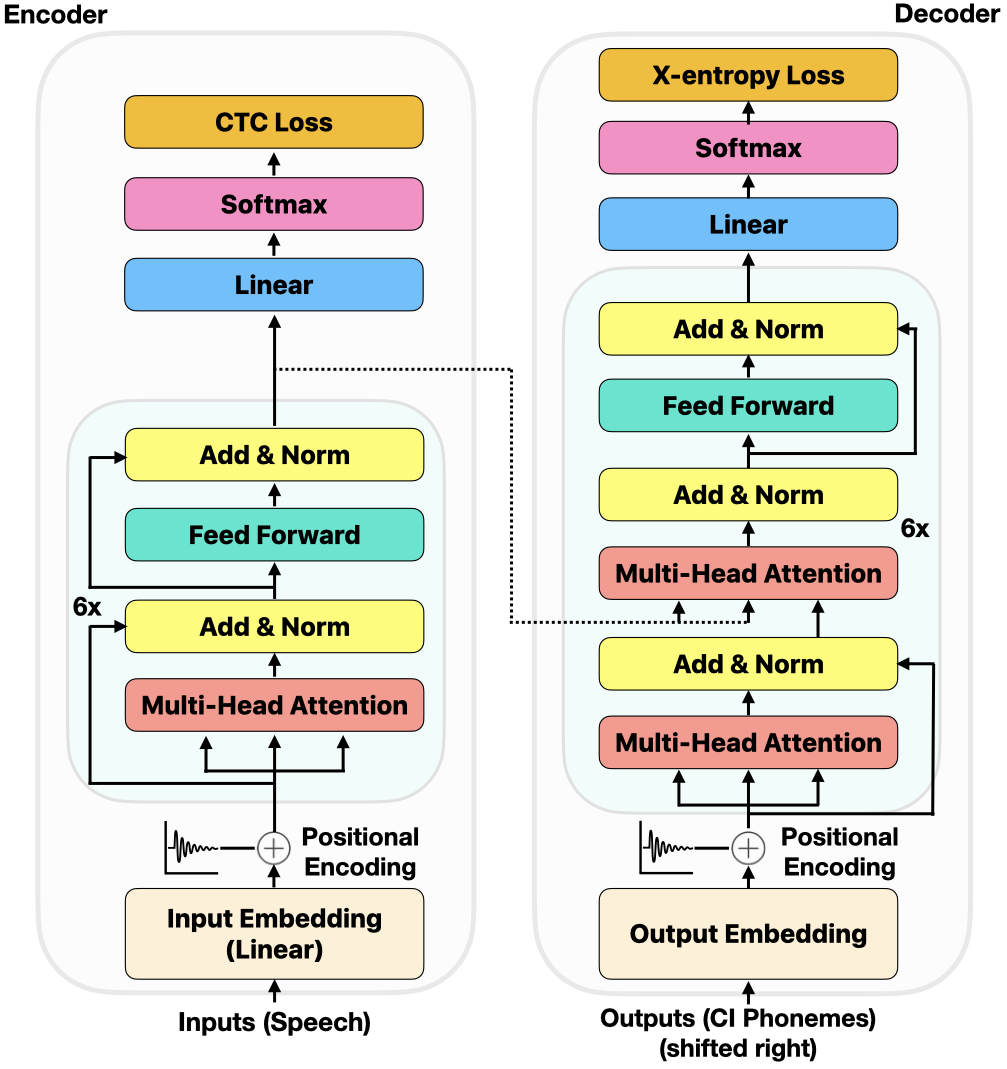}
\caption{Proposed network architecture. The encoder comprises a stack of self-attention layers and is trained by minimizing the CTC loss (left). The dashed arrow shows an optional decoder that can be added during training (right). Note that for all configurations, only the encoder branch (left) is used during inference.}
\vspace{-5mm}
\label{fig:model_arch}
\end{figure}

Recently, there have been several studies \cite{watanabe-hybrid,yuan2018improved,Xiao-isclp2019} that suggest that the accuracy of a sequence-to-sequence architecture (an encoder and an autoregressive decoder with cross-attention) for speech recognition \cite{chan2016listen} can be improved by jointly minimizing the cross-entropy loss on the decoder \emph{and} the CTC loss acting on the outputs of the encoder. The intuition is that these two losses regularize each other resulting in faster training convergence and more accurate models. We experiment with such an architecture (dashed arrow in Figure \ref{fig:model_arch}). At inference, we only use the encoder branch of the network. So effectively the decoder trained to minimize the cross-entropy loss is acting as an additional regularization term added to the network described in Section \ref{sec:attention_ctc}. Since we do not use the decoder at inference, the number of parameters of this model and the one describe above remain the same. The architecture of the decoder is depicted in Figure \ref{fig:model_arch}. We use a stack of 6 layers in the decoder keeping the parameters of the self-attention and the feed-forward layers exactly the same as the encoder. We linearly combine the cross-entropy loss and the CTC loss for every utterance with unity coefficients. We refer to this configuration as \emph{Transformer Encoder} in the rest of the paper. 

%\section{Hybrid Transformer/CTC network}
%
%\input{input/tfencoder}

\section{Multi-task Learning}

\label{sec:mtl}

The model architectures outlined in Section \ref{sec:background} are all monophone AMs that are trained to minimize the CTC loss or a combination of the CTC loss and the cross-entropy loss for the models in Section \ref{sec:attention_ctc_decoder}. As argued in \cite{sigtia2020mtl}, this training objective does not match the final objective we care about, which is to discriminate between examples of true triggers and phonetically similar acoustic segments. Previously we showed that we can achieve significant performance improvements in trigger detection by adding a relatively small amount of trigger phrase specific \emph{discriminative} data and finetuning a pre-trained phonetic AM to minimize the CTC loss and the discriminative loss simultaneously \cite{sigtia2020mtl}. We apply this idea to the models described in Section \ref{sec:background}. For each of the model architectures, we take the encoder branch of the model and add an additional output layer (affine transformation + softmax non-linearity) with 2 output units at the end of the encoder network. One unit corresponds to the trigger phrase, while the other unit corresponds to the negative class. The objective for the discriminative branch is as follows: for positive examples we minimize the loss $C = -\max_{t} \log y^{P}_{t}$, where $y^{P}_t$ is the network output at time $t$ for the positive class. This loss function encourages the network to yield a high score \emph{independent of the temporal position}, note that this is only useful for networks that read the entire input at once. For negative examples, the loss function is $C = -\sum_{t} \log y^{N}_{t}$, where $y^{N}_{t}$ is the network output for the negative class at time $t$. This loss forces the network to output a high score for the negative class \emph{at every frame}.

%we add an additional output layer (affine transformation + softmax non-linearity) with 2 output units, to the outputs of the encoder network. One unit corresponds to the trigger phrase while the other unit corresponds to the blank symbol used by the CTC training objective. Then to each of the objective functions outlined before, we add an additional discriminative loss term as described in \cite{sigtia2020mtl}. We simply sum up the objective functions with unity coefficients. Note that when performing MTL, we do not use the decoder branch. We only minimize the 2 CTC losses, one for the monophone AM branch as before and one for the discriminative branch \cite{sigtia2020mtl}.

\section{Model Training}

\label{sec:setup}

\subsection{Monophone AM Training Data}

We follow a similar pipeline as \cite{sigtia2020mtl} for preparing the training data for the monophone AMs described in Section \ref{sec:background}. We start with a \emph{clean} dataset with about 2700 hours of transcribed audio. These examples are recorded on mobile phones and therefore are assumed to be \emph{near field}. For each utterance in the dataset, we augment it by convolving the audio with a room impulse response (RIR) that is randomly selected from a set of 3000 RIRs. This process yields a reverberated copy of the original dataset. Next, we collect over 400,000 examples of echo residuals from various devices playing music, podcasts and text-to-speech at varying volumes \cite{MLBlogFrontEnd}. We then mix each example in the reverberated dataset with a randomly selected echo residual from the corpus, resulting in over 8700 hours of transcribed and augmented training data for the AMs. We pick training examples that explicitly \emph{do not} contain the trigger phrase in order to avoid biasing the AMs. 

\subsection{Multi-Task Training Data}

We use the same dataset as described in \cite{sigtia2020mtl} for MTL experiments: 40,000 examples that false trigger the baseline system and another 140,000 examples of true trigger phrases. We run a first-pass DNN-HMM detector on the audio to obtain trigger start and end boundaries for each utterance. We then extract only these segments from each utterance, which results in 90 hours of audio. For MTL experiments, we concatenate the AM training dataset and the discriminative dataset and randomly sample mini-batches from the combined dataset. 

\subsection{Training Hyper-parameters}

We use exactly the same hyper-parameters for training all the models. We use mini-batches of 32 utterances per GPU with 16 GPUs in parallel and an initial learning rate of 5e-5. We use the Adam optimizer \cite{kingma2014adam} and synchronous gradient updates. We stop training if the validation loss does not improve after 8 consecutive training epochs.

\section{Evaluation}

\label{sec:results}

\subsection{Datasets}

We use the same datasets for evaluation as described in \cite{sigtia2020mtl} with no changes and therefore the results are directly comparable. Both datasets are internally collected specifically for the purpose of evaluating voice trigger models. Both datasets are collected using smart speakers in a variety of environments and conditions to simulate real-world usage. The first \emph{structured} dataset contains utterances from 100 participants, approximately evenly divided between male and female. Each subject speaks a series of prompted voice commands, where each command is preceded by the trigger phrase. The recordings are made in 4 different acoustic settings: quiet room, external noise from TV or kitchen appliances, music playing from the recording device at medium volume and finally music playback from the device at loud volume. The final condition is the most difficult since the input signals have a considerable amplitude of residual noise. We collect 13,000 such \emph{positive} utterances. These examples allow us to measure the number of false rejections (FRs) made by the system. We also use a set of 2000 hours of audio recordings comprising Podcasts, audiobooks, TV playback etc. This audio \emph{does not} contain the trigger phrase and acts as a \emph{negative} set that allows us to estimate the number of false alarms (FAs) per hour of active audio. 

We also conduct a second \emph{unstructured} data collection at home by our employees. We ask 42 participants to use a smart speaker daily for 2 weeks. We enable extra logging and review by the users in order to allow them to choose which recordings they want to delete. This setup allows us to collect data that represents more spontaneous device usage (non-stationary sources, non-stationary noise, children's speech, overlapping speech etc.). Continuous recording for 2 weeks on a device is not possible therefore we use a first-pass DNN-HMM system \cite{MLBlogHS} with a low-threshold that detects audio segments phonetically similar to the trigger phrase, allowing us to measure (almost) unbiased false-reject rates for realistic in-home usage. (With customer data, the audio sent to the server has already triggered the device, therefore making it impossible to measure false reject rates.)

\subsection{Results}

% \begin{figure*}[htp!]
%   \centering
%   \subfigure[random caption 1]{\includegraphics[scale=0.45]{figures/DET-Curve_Checker_BiLSTM_SA_TF_Encoder_MTL_HomePod_EDC}\label{fig:det_curve_checker_bilstm}}
%   \subfigure[random caption 2]{\includegraphics[scale=0.45]{figures/DET-Curve_Checker_BiLSTM_SA_TF_Encoder_MTL_HomePod_THK}\label{fig:det_curve_checker_bilstm}}
% \end{figure*}

\begin{figure*}[htpb!]
\centering
\begin{minipage}[b]{.4\textwidth}
\includegraphics[scale=0.45]{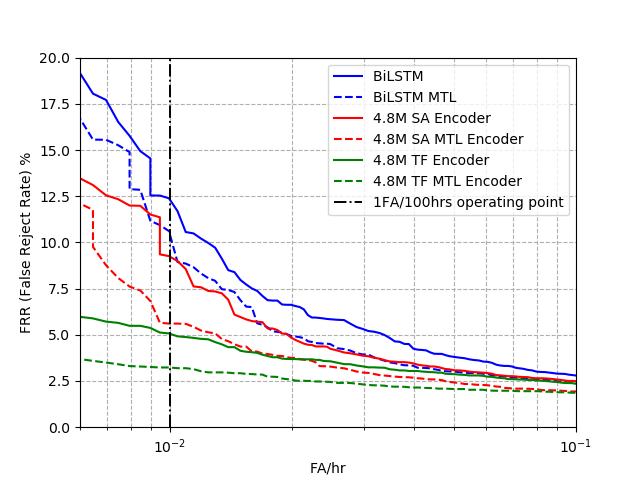}
\caption{DET curves for structured evaluation set}\label{fig:det_curve_checker_edc}
\end{minipage}\qquad
\begin{minipage}[b]{.4\textwidth}
\includegraphics[scale=0.45]{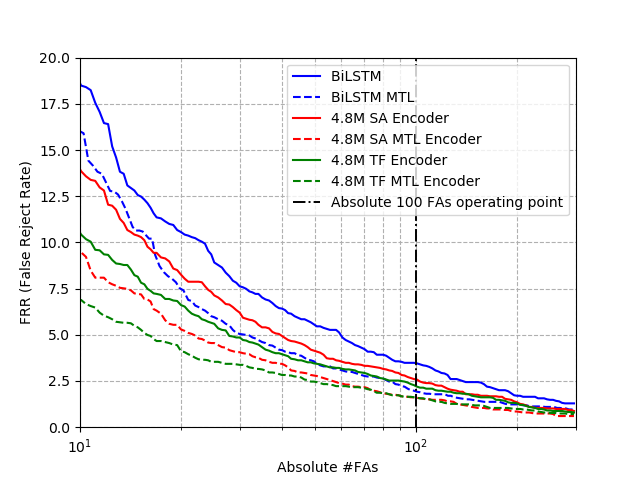}
\caption{DET curves for take-home evaluation set}\label{fig:det_curve_checker_thk}
\end{minipage}
\vspace{-5mm}
\end{figure*}

Figure \ref{fig:det_curve_checker_edc} presents modified detection error tradeoff (DET) curves for all the models evaluated on the structured evaluation dataset. The (log) X-axis represents the number of false alarms (FAs) per hour of active audio while the Y-axis represents the proportion of false rejects (FRs) by the system (lower is better). Solid curves represent the baseline models trained only on the large AM training dataset, while the dashed curves represent the MTL versions of these models trained on the dataset described in Section \ref{sec:setup}. From Figure \ref{fig:det_curve_checker_edc}, note that the model with self-attention layers trained with the CTC objective function (red) yields much better accuracy than the baseline BiLSTM model (blue). In fact, this model yields similar accuracies as the MTL version of the BiLSTM model (dashed red). This is a significant result as the new model was trained without any additional discriminative data and therefore the improvements can be attributed entirely to the change of layer architecture. Additionally, the self-attention model has 10\% fewer parameters than the baseline BiLSTM model. Next, the model with self-attention layers trained with a decoder loss (green) yields more than 50\% relative improvement over the BiLSTM baseline (Table 1). This model is trained only on the AM training set but it is more accurate than the MTL versions of the both the baseline BiLSTM (dashed blue) and self-attention + CTC model (dashed red). This result suggests that adding the decoder as an additional loss results in significant improvements without adding any extra parameters to the model (since we only use the encoder part of the network). Again, this result is practically useful since collecting a dataset of difficult negative examples is challenging and is not available for many languages and these new models yield better results than the MTL versions of the baseline \emph{without} requiring any discriminative training data. Finally for all architectures, the MTL versions of the models (dashed curves) always yield significant improvements over the baselines. Table 1 presents FR rates at an operating point of 1 FA per 100 hours for the different models. 

%\vspace{1mm}
%\begin{table}[htpb]
%\small
%\centering
%\begin{tabular}{lcc}
%\hline 
%Model & False Reject & Relative \\
%Architecture & Rate (\%) & Impr (\%) \\
%\hline 
%BiLSTM & 11.7 & 0\\
%BiLSTM MTL & 8.9 & 23.9\\ 
%Self Attention Encoder & 8.9 & 23.9 \\
%Self Attention Encoder MTL & 5.6 & 52.1 \\
%Transformer Encoder & 4.9 & 58.1 \\
%Transformer Encoder MTL & 4.1 & 64.9\\
%\hline
%\end{tabular}
%\caption{False Reject Rate at an operating point of 1 FA/100 hrs for structured evaluation set}
%\label{tab:frr}
%\end{table}

\vspace{1mm}
\begin{table}[tpb!]
\small
\centering
\begin{tabular}{lcc}
\hline 
Model & Phonetic Training & MTL Training\\
Architecture & FRR (\%) & FRR (\%)\\
\hline 
BiLSTM & 11.7 & 8.9\\
Self Attention Encoder & 8.9 & 5.6 \\
\textbf{Transformer Encoder} & \textbf{4.9} & \textbf{3.2} \\
\hline
\end{tabular}
\caption{False Reject Rate (FRR) at an operating point of 1 FA/100 hrs for structured evaluation set}
\vspace{-5mm}
\label{tab:frr}
\end{table}

\vspace{1mm}
\begin{table}[tpb!]
\small
\centering
\begin{tabular}{lcc}
\hline 
Model & Phonetic Training & MTL Training \\
Architecture & FRR (\%) & FRR (\%) \\
\hline 
BiLSTM & 3.4 & 1.9\\
Self Attention Encoder & 2.6 & 1.6 \\
\textbf{Transformer Encoder} & \textbf{2.2} & \textbf{1.6}\\
\hline
\end{tabular}
\caption{False Reject Rate (FRR) at an operating point of 100 FAs on unstructured take home evaluation set}
\vspace{-5mm}
\label{tab:frr}
\end{table}

Figure \ref{fig:det_curve_checker_thk} presents DET curves for the unstructured evaluation dataset. We observe a similar trend as above. The self-attention network trained with the CTC loss (red) improves over the baseline BiLSTM network (blue). Next, the self-attention network trained with both the CTC loss and the additional decoder yields further improvements (green), notably yielding better accuracies than the MTL version of the BiLSTM baseline (dashed blue). Finally, the MTL versions of both the self-attention networks yield significant improvements over all the baselines. Table 2 presents the FR rate at 100 FAs for the different architectures considered.

\subsection{Hardware Efficiency}

We compare the on-device inference times for the baseline BiLSTM model (5.5M params) and the Transformer Encoder (4.8M params) in Table \ref{tab:inference_time}. At inference, Self Attention Encoder and Transformer Encoder perform exactly the same computation. Inference was performed on 1.8 seconds of audio sampled at 16KHz. The compute platform was a 2019 smart phone with fixed CPU core and frequency. Compared to baseline 8-bit 5.5M BiLSTM model, using an 8-bit quantized 4.8M Transformer Encoder model, we obtain an improvement of 70\% in inference speed. The huge improvements in runtime can be attributed to the self-attention layers not having sequential dependencies and being more parallelizable than BiLSTMs. Hence self-attention layers can make better use of specialized hardware on a modern processor for accelerating matrix computations. Table \ref{tab:inference_devices} also shows that the compute speed advantage of the Transformer Encoder model is less on a variety of older platforms. In addition to faster inference, Table 5 shows that the Transformer Encoder models are significantly faster ($\sim$ 5x) to train.

\begin{table}[tpb]
\small
\centering
\begin{tabular}{lccc}
\hline 
%& \multicolumn{2}{c}{Per frame inference time (in ms) on device}\\
Metric & BiLSTM & \textbf{Transformer} & Impr.\\
       & (5.5M) & \textbf{Encoder (4.8 M)} & (\%)\\ 
\hline 
Network runtime  & 95 ms & \textbf{28 ms} & 70.5\\
Network memory  & 5.6 mb & \textbf{4.7 mb} & 16.1\\ 
Network runtime & 5.7 mb & \textbf{2.7 mb} & 52.6\\
~~memory & & & \\

\hline
\end{tabular}
\caption{Network runtime and memory usage on 2019 smart phone}
\vspace{-5mm}
\label{tab:inference_time}
\end{table}

\begin{table}[tpb]
\small
\centering
\begin{tabular}{lcc}
\hline 
Platform & \% Improvement in Network Runtime \\
\hline
Smart Speaker & 20.8\% \\
Smart Watch & 30.5\% \\
2015 Smart Phone & 17.8\% \\
\hline
\end{tabular}
\caption{Network runtime improvements of 4.8M TF encoder over BiLSTM on various platforms}
\vspace{-5mm}
\label{tab:inference_devices}
\end{table}

%Model Architecture & epochs & Training Time (in minutes) \\
%\vspace{1mm}
%\begin{table}[h!]
%\small
%\centering
%\begin{tabular}{lcc}
%\hline 
%Model & epochs & Utterances\\
%Architecture & pp & per second \\
%\hline 
%BiLSTM &  &  & 10080 \\
%Transformer Encoder &  &  & 1862\\
%Self Attention Encoder &  &  & 1304\\
%Transformer &  &  & 1502 \\
%\hline
%\end{tabular}
%\caption{Mean Training Time with 16 GPUs}
%\vspace{-5mm}
%\label{tab:training_time}
%\end{table}

\begin{table}[tpb!]
\small
\centering
\begin{tabular}{lccc}
\hline 
Model & Utterances & Epochs & Train Time\\
Architecture & per second &  & (minutes)\\
\hline 
BiLSTM  & 210 & 115 & 10080\\
\textbf{Self Attention Encoder} & \textbf{1121} & \textbf{77} & \textbf{1304}\\
Transformer Encoder & 788 & 75 & 1862\\

%Transformer & 1080 & 85 & 1502\\

\hline
\end{tabular}
\caption{Average AM training time statistics over 5 runs.}
\vspace{-5mm}
\label{tab:training_time}
\end{table}

\section{Conclusions}

\label{sec:conclusion}
In this work we study the problem of designing a hardware efficient voice trigger detection system.
We start with a BiLSTM network trained to minimize the CTC loss.
We explore replacing the BiLSTM layers with self attention layers and show improvements in accuracy and inference times.
We propose to regularize the training process by adding an auto regressive transformer decoder with a cross entropy loss and show significant improvements in accuracy.
We then improve the results further by using
MTL on the encoder outputs, with the additional task being a true trigger/false trigger classifier.
We show that compared to baseline BiLSTM approach, the hybrid transformer/CTC setup significantly improves the FRR by $\sim$60\% for a given FAR (1 FA/100 hrs) with 10 \% fewer model parameters. Additionally, the proposed approach reduces the on-device inference
time by 70\% and is $\sim$5X faster to train.

%\section{Acknowledgements}
%\newpage
\bibliographystyle{IEEEtran}

\bibliography{mybib}

\end{document}